\documentclass[%
 reprint,
superscriptaddress,
 amsmath,amssymb,
 aps,
]{revtex4-2}

\usepackage{graphicx}% Include figure files
\usepackage[dvipsnames]{xcolor}
\usepackage{todonotes}
\usepackage{dcolumn}% Align table columns on decimal point
\usepackage{bm}% bold math
\usepackage{braket}
\usepackage[markup=underlined]{changes}
\newcommand{\red}[1]{\textcolor{black}{#1}}

\begin{document}
\preprint{APS/123-QED}

\title{\red{Non-linear} correlations underlie linear response and causality}% Force line breaks with \\

\author{Gabriele Di Antonio}
\email{gabriele.diantonio@cref.it}
\affiliation{Natl. Center for Radiation Protection and Computational Physics, Istituto Superiore di Sanità, 00161 Rome, Italy}
\affiliation{Research Center ``Enrico Fermi'', 00184 Rome, Italy}
\affiliation{Engineering department, ``Roma Tre'' University of Rome, 00146 Rome, Italy}

\author{Gianni Valerio Vinci}
\email{gianni.vinci@iss.it}
\affiliation{Natl. Center for Radiation Protection and Computational Physics, Istituto Superiore di Sanità, 00161 Rome, Italy}

\date{\today}% It is always \today, today,

\begin{abstract}

The inference of causal relationships among observed variables is a pivotal, longstanding problem in the scientific community. An intuitive method for quantifying these causal links involves examining the response of one variable to perturbations in another. The fluctuation-dissipation theorem elegantly connects this response to the correlation functions of the unperturbed system, thereby bridging the concepts of causality and correlation. However, this relationship becomes intricate in nonlinear systems, where knowledge of the invariant measure is required but elusive, especially in high-dimensional spaces.
In this study, we establish a novel \red{equation that links} the Koopman operator of nonlinear stochastic systems and the response function. This connection provides an alternative method for computing the response function using generalized correlation functions, even when the invariant measure is unknown. We validate our theoretical framework by applying it to \red{several nonlinear systems}, showing convergence and consistency \red{for increasing number of measured observables and samples.}
Finally, we discuss a significant interplay between the resulting causal network and the relevant time scales of the system.

\end{abstract}

\maketitle

%\tableofcontents
Inference of causal relations between measured quantities is a central and old open problem in Science. The network of causal links between available degrees of freedom is essential to build effective models of reality. As the old adage goes ``correlations do not imply causation'', however a proper use of statistical information can go a long way in predicting causal relationships. For example in their seminal work Wiener \cite{wiener1956theory} and Granger \cite{granger1969investigating} proposed that  $Y$ is causally related to $X$ if the forecasting of $X$ based on it's previous values increase when the information of $Y$ is added. An alternative approach, also based on forecasting, is rooted in the embedding theory of dynamical systems pioneered by Takens \cite{kantz2003nonlinear}. Despite the fact that both approaches have intrinsic limitations, they have been massively adopted in virtually all fields of Science \cite{runge2023causal,setH2015granger,vinci2018economic,javarone2023disorder}. There is however another possible definition of causality commonly accepted in physics. We say that $X$ is causally related to $Y$ if a small perturbation on the latter has a significant effect on the former. The quantity that measure this effect is commonly referred to as the response function $R_{Y \to X}(t)$. While estimating causation via intervention is certainly a good idea, practical limitations often hinder our ability to clearly perturb certain degrees of freedom and observe the system's response. This issue can be bypassed thanks to fluctuation dissipation theorem \cite{marconi2008fluctuation} that relates the response to generalized correlation functions of the unperturbed system. Building on this concept, Baldovin et al. \cite{baldovin2020understanding} have proposed to measure causality as time integral of $R_{Y \to X}(t)$ to account for potential time-lagged causal effects. When the system is linear the response is proportional to the standard cross correlations functions between all possible pairs of degrees of freedom, see also \cite{baldovin2022extracting,baldovin2021handy,lucente2022inference}. For general non-linear systems the fluctuation-dissipation theorem can still be applied but requires the knowledge of the functional form of the invariant probability density functions $p_s(\bm x)$ \cite{marconi2008fluctuation} which is non trivial to estimate for high-dimensional problems. In this paper we apply the Koopman formalism \cite{rowley2009spectral,klus2020data} and show that by lifting the system to higher dimensions, introducing extra degrees of freedom, a simple and general relationship between response and correlations among the extended set of variables can be recovered, analogous to what occurs in linear systems. \red{While here we focus mainly on causal detection we note that the connection between Koopman operator theory and response functions has recently been investigated also in the following preprint \cite{zagli2024bridging}. \\
We begin by establishing a novel relation between the correlations computed in the Koopman operator basis and the response function. We first apply our findings to an instructive non-linear example where the high dimensional lifting can be carried out analytically.
We test our approach on various dynamical systems and we explore the convergence and constrains for increasing dimensionality. Finally utilizing the computed $R(t)$ we use our results to infer causality as proposed in \cite{baldovin2020understanding}.}

\begin{figure*}[ht!]
    \centering
    \includegraphics[width=180mm]{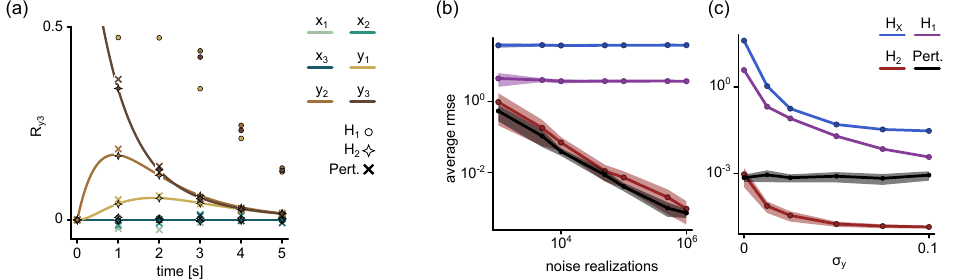}
    \caption{(a), Various estimations of the response function in relation to the perturbing variable $y_3$ for a six-dimensional example system as described in \eqref{eq:toyexample}. The solid line represents the ground truth, while the individual points denote the definitions of perturbation and the fluctuation correlation formula presented in \eqref{eq:MainResult}, specifically for the first-order $\mathrm{H}_1$ and second-order $\mathrm{H}_2$ Hermite polynomial observables. Averages over $10^6$ experiment realizations for systems parameters $\sigma_x=0.1$,  $\sigma_y=0.025$.
    (b-c), The average reconstruction error of response curves across all dimensions and time steps, estimated using root mean square error, for the same system depicted in (a).
    Curves are estimated by perturbation limit and from \eqref{eq:MainResult} using observables $\psi_k(\mathbf x) = x_k$, named $\mathrm H_x$, and Hermite basis $\mathrm H_1$, $\mathrm H_2$.
    This analysis considers respectively (b) varying numbers of process realizations ($\sigma_y=0.0$) and (c) different levels of noise $\sigma_y$ ($10^6$ experiments). The shaded ribbon area illustrates the standard deviation resulting from $10$ complete analysis repetitions. The time resolution is set at $\mathrm{dt}=0.01 s$, perturbation amplitude $\epsilon=0.01$, with stochastic Heun method for integration.
    }
    \label{fig:1}
\end{figure*}

\section*{Koopman operator and generalized linear response} In the following we will consider the generic stochastic markovian dynamical systems of the form:
\begin{equation}\label{eq:stocSystem}
    d\mathbf{X}=a(\mathbf{X})dt +b(\mathbf{X})d\mathbf{W}
\end{equation}
where $\mathbf{X} \in R^{N}$ is an $N$ dimensional stochastic process $a(\mathbf{x}) \in R^{N}$ is the drift term and $b(\mathbf{X}) \in R^{N \times S}$ is the diffusion term while $d\mathbf{W} \in R^{S}$ are independent Wiener noise sources \cite{gardiner2009stochastic}. We also assume natural boundary conditions on a domain $\mathcal{D}$. For these systems, the Koopman operator $\mathcal{K}$ governs the time evolution of observables $f$:
\begin{equation}\label{eq:KoopmanDef}
    \mathcal{K}^t f (\mathbf{X}_0)= \mathrm E[f(\mathbf{X}_t)|\mathbf{X}_0] 
\end{equation}
where $\mathrm E$ indicates the expectation value, conditioned to $\mathbf{X}(t=0)=\mathbf{X}_0$. Note that for systems like (\ref{eq:stocSystem}) the associated infinitesimal generator $\mathcal{L}f=\lim_{t\to 0} \frac{\mathcal{K}^{t}f -f}{t}$ is the backward Kolmogorov operator \cite{klus2020data,risken1996fokker} whose eigenvalues $\lambda_n$ dictates the relevant time scales of the dynamics and are also involved in the first passage time problem \cite{vinci2024escape}. Since it is in general difficult to find the eigenfunctions $\mathcal{L}\phi_n(\mathbf{x})=\lambda_n \phi_n(\mathbf{x})$ we can resort to a pseudospectral approach \cite{shizgal2015spectral} which in the Koopman formalism is tipycally refered to as Extended Dynamic Mode Decomposition \cite{williams2015data}. 
Consider a set of $d$ observables, or basis functions, $\bm \psi(\bm x)=\{ \psi_k(\bm x) \}_{k=1}^{d}$ that ideally span a subspace containing the state of the system at all times. We chose the basis to be orthogonal, and normalized, with respect to the inner product defined by the measure $w(\bm x)$ i.e. $\int_{\mathcal{D}} \psi_i(\bm x)\psi_j(\bm x) w(\bm x) d\bm x=\delta_{ij} $. Then the matrix $\mathbf K (t)= e^{\mathbf L t}$ where $\mathrm L_{nm}=\int_{\mathcal{D}}\psi_{n}(\bm x)\mathcal{L}\psi_{m}(\bm x) d\bm x$, approximate the action of the Koopman operator i.e.:
\begin{equation}\label{eq:Konpsi}
    \mathrm{E}[\bm \psi (\mathbf X_t) | \mathbf X_0] \approx \mathbf K(t) \bm \psi(\mathbf X_0)
\end{equation}

While for $d \to \infty$ convergence is guaranteed and $ \mathbf L$ becomes iso-spectral to $\mathcal{L}$, in practical application controlling spectral pollution due to finite $d$ is still an open problem which we will not address here \cite{colbrook2023residual,colbrook2019compute}. Notably if we chose as basis function the eigenfunctions of $\mathcal{L}$, $\mathrm K_{nm}(t)=e^{\lambda_n t} \delta_{nm}$. 
In the following we will assume the existence of a matrix $\bm \alpha$ that relates physical degree of freedom $\bm x$ with the observables $\bm \psi$ i.e.: $\bm x \approx \bm \alpha \bm \psi(\bm x)$. Then, expected values can be expressed as follow:

\begin{equation}\label{eq:exceptedvalues}
\mathrm{E}[ \mathbf X_t | \mathbf X_0] \approx \bm \alpha \mathbf K(t) \bm \psi (\mathbf X_0)
\end{equation}
Here we will treat $\bm{\psi}$ as independent degrees of freedom, which we will refer to as virtual to distinguish them from the physical degree of freedom $\bm x$, similarly to the auxiliary response field introduced in statistical physics \cite{tauber2014critical}.
For the response function we follow  \cite{baldovin2020understanding} and we will indicate with $\{\cdot\}_p$ the system perturbed in the initial condition. The virtual response seen on the physical variable $x_j$ due to a perturbation in the virtual variable $\psi_i(\mathbf X_0) \rightarrow \psi_i(\mathbf X_0) + \epsilon$  
is defined as follow
\begin{equation}\label{eq:responseDef}
        \tilde{\mathrm R}_{\psi_i \rightarrow x_j}(t) = \lim_{\epsilon \to 0} \frac{\langle  \mathrm X_j(t)\rangle _p -\langle  \mathrm X_j(t)\rangle}{\epsilon}
\end{equation}

where the brackets include the average with respect to the distribution of the initial conditions \footnote[1]{ $\braket{f(x)}= \int_{\mathcal{D}} f(x)  p(x,t|x_0) p_s(x_0) dx dx_0$}. From \eqref{eq:exceptedvalues} follows the relation of the virtual response to the Koopman matrix:
\begin{equation}\label{eq:VirturalResponse}
    \tilde{\mathrm R}_{\psi_i \rightarrow x_j}(t) \approx [\bm \alpha \mathbf K (t)]_{ji}
\end{equation}
\red{Clearly eq. (\ref{eq:VirturalResponse}) cannot be used in practice because there is no way to perturb such virtual degrees of freedom without changing $\mathbf{X}(t)$, however as we will show this perturbation is not needed.}
In the same way we can look at the effect of perturbing $x_i(0) \rightarrow x_i(0) + \epsilon$, and finally uncover the relation between the standard Response function and \eqref{eq:VirturalResponse}:
\begin{equation}\label{eq:CentralResult}
  \mathrm R_{x_i\rightarrow x_j}(t) \approx  \sum_{m=1}^d \left \langle \left. \frac{\partial \psi_m(\mathbf x)}{\partial x_i} \right |_{\mathbf X_0} \right \rangle \tilde{\mathrm R}_{\psi_m \rightarrow x_j}(t)
\end{equation}
The expansion in the dimensionality of the system, by including the virtual degrees of freedom has a linearizing effect which is also the major appeal of the Koopman formalism \cite{brunton2022modern}. In turns, this suggests that we can relate the virtual response to correlations of the unperturbed system. In fact, applying $\bm \alpha$ to \eqref{eq:exceptedvalues} and using \eqref{eq:VirturalResponse} we find:
\begin{equation*}
    \mathrm{E}[\mathbf X_t | \mathbf X_0] \bm \psi(\mathbf X_0)^T \approx \mathbf{\tilde{R}}_{\psi \rightarrow x}(t) \bm \psi(\mathbf X_0) \bm \psi(\mathbf X_0)^T
\end{equation*}

By averaging over the initial condition we find a fluctuation-dissipation type relation:
\begin{equation}
    \mathbf{\tilde{R}}_{\psi \rightarrow x}(t) \approx \mathbf C_{x\psi}(t) \mathbf C_{\psi \psi}(0)^{-1}
    \label{eq:virtualrespC}
\end{equation}
where $\mathbf C_{fg}(t) = \braket{\bm f(\mathbf X_t) \bm g (\mathbf X_0)^T}$  are covariance matrices in case of null averages. It is worth noting that \eqref{eq:virtualrespC} does not stand on the semi-group condition, providing an autonomous estimate of $\mathbf K (t)$ for every time point, which enhances the robustness of the method by mitigating error propagation.
Finally, equation \eqref{eq:virtualrespC} can be used to find the relation between the linear response of the system and unperturbed correlations with respect to virtual degrees of freedom, \red{one of the main results of this paper}:
\begin{equation}\label{eq:MainResult}
\textbf{R}(t) \approx \mathbf{C_{x\psi}}(t) \mathbf{C_{\psi \psi}}(0)^{-1} \braket{\nabla \bm \psi}
\end{equation}
where $\braket{\nabla \bm \psi}_{ij}=\left \langle \left. \frac{\partial \psi_i(\mathbf x)}{\partial x_j} \right |_{\mathbf X_0} \right \rangle$ and $\mathrm{R}_{ij} = \mathrm R_{x_j\rightarrow x_i}$.
Notably the proportionality of the linear response of systems like \eqref{eq:stocSystem} to non-standard correlations can be inferred starting from the fact that under rather general assumption \cite{marconi2008fluctuation}
\begin{equation}\label{eq:GeneralizedFD}
     \mathrm R_{x_i\rightarrow x_j}(t) = - \left \langle \mathrm X^{(j)}_t  \left. \frac{\partial \log[p_s(\bm x)]}{\partial x^{(i)}} \right |_{\mathbf X_0} \right \rangle
\end{equation}
In the case in which the considered invariant subspace contains the functions  $\bm \nabla \log[p_s(\bm x)]$ (i.e. $\exists \, \bm \beta$; $\bm \nabla \log[p_s(\bm x)] \approx \bm \psi(\bm x) \bm \beta^T$), we can rewrite \eqref{eq:GeneralizedFD} as:
\begin{equation}
    \mathbf R(t) \approx \mathbf{C_{x\psi}}(t) \bm \beta^T
    \label{eq:RespDistr}
\end{equation}
Contrary to \eqref{eq:MainResult} where we need the stationary expectation values of the virtual degree of freedom that can be sampled via Monte-Carlo methods, to apply \eqref{eq:GeneralizedFD} the knowledge of functional form of $p_s (\bm x)$ is required which is not always easy to get expecially for high-dimensional systems. The relation between \eqref{eq:MainResult} and \eqref{eq:RespDistr} is more clear if we chose as weight of the inner product the stationary distribution i.e. $w(\bm x)= p_s(\bm x)$. In this case then, the normalization condition implies that $ \mathbf{ C_{\psi \psi}}(0) =\mathcal{I}$ is the identity matrix, while from the definition of $\bm \beta$ and integration by parts follows that: 
\begin{equation*}
\begin{split}
      \beta_{ij} &=-\int_{\mathcal{D}} \psi_{i}(\bm x) \frac{ \partial_j p_s (\bm x)}{p_s(\bm x)} w(\bm x) d\bm x \\ 
      &=\int_{\mathcal{D}} \frac{\partial \psi_i (\bm x)}{\partial x_j} p_s (\bm x)  d\bm x = \braket{\nabla \bm \psi}_{ij}
\end{split}
\end{equation*}
hence \eqref{eq:RespDistr} and \eqref{eq:MainResult} are, in this case, equivalent.

\section*{Linear response in non-linear systems}
For system with linear drift and constant diffusion, i.e. a multidimensional Ornstein–Uhlenbeck process \cite{gardiner2009stochastic}, the choice $\psi_k(\bm x) = x_k$ is sufficient to fully capture the response. In fact in this case the stationary distribution $p_s(\bm x)$ is gaussian and from \eqref{eq:GeneralizedFD} it follows : $\textbf{R}(t) = \mathbf{C_{x x}}(t) \mathbf{C_{xx}}(0)^{-1}$. 
\begin{figure*}
    \centering
    \includegraphics[width=170mm]{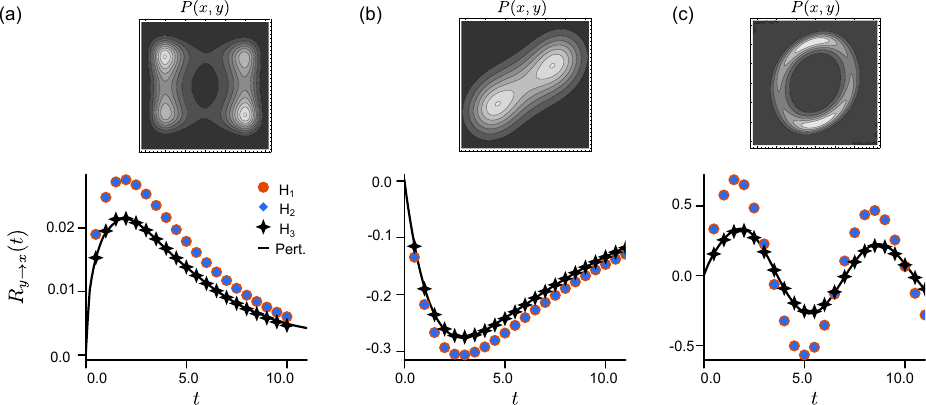}
    \caption{ \red{In the top row, we show the stationary distribution for system (\ref{eq:GradSystem}) in panel (a), (\ref{eq:TanhSystem}) in panel (b), and (\ref{eq:Stuart-Landau}) in panel (c). In the bottom row, the response functions (in black) are computed from the time-dependent solution of the Fokker-Planck equation, while the scatter points represent the result of equation (\ref{eq:MainResult}) for different basis functions $H_1$, $H_2$, and $H_3$, as shown in Fig. \ref{fig:1}. The parameters are $\alpha=0.1$, $a=1$, $b=-0.3$, $J_{11}=0.9$, $J_{12}=0.5$, $J_{21}=0.3$, $J_{22}=0.6$, and $\sigma=0.3$.}}
    \label{fig:2DCases}
\end{figure*}
Recently in \cite{baldovin2020understanding} it has been showed that the fluctuation-dissipation formula for linear systems can sometimes yield satisfactory results even in the context of nonlinear dynamics. This is possible only when the stationary distribution is well approximated by a gaussian for the reasons mentioned above. On the other hand our result \eqref{eq:MainResult} is the natural generalization that applies for more broad cases paying the cost of extending ``virtually'' the dimensionality of the systems.
In the Koopman literature, the proposed generalization parallels the extension of the Dynamic Mode Decomposition in \cite{williams2015data}. We note however that analogous conclusions can be achieved without invoking the Koopman formalism at all. To do so we can use a generalization of the regression theorem, an important result of stochastic calculus \cite{gardiner2009stochastic} that has found several applications in quantum mechanics \cite{gardiner2004quantum}. The theorem is generally presented for stochastic systems with linear drift and constant diffusion but it can be generalized also to multiplicative noise as long as the drift component remains linear. For a $d$- dimensional process $\mathbf{X}$ such that:
$$d \mathbf{ X} = \mathbf A \mathbf{ X}  dt + b(\mathbf{ X}) d \mathbf{W}$$
one can show that interdependently of the specific choice of $b(\mathbf{ X})$, the time dependent evolution of the vector of observables $\mathbf{m}(t)=\{E[X_{i,t}|\mathbf{X}_0]   \}_{i=1}^{d}$ is simply given by: 
\begin{equation}\label{eq:RegresionTh}
\mathbf{m}(t)=e^{\mathbf{A} t}\mathbf{m}(0)    
\end{equation}
Now we go back to the general system (\ref{eq:stocSystem}) and suppose we have found a collection of virtual degrees of freedom, i.e. functions $Z_n=f(\mathbf{X})$, such that the extended system with $\tilde{\mathbf{X}}=(\mathbf{X}...\mathbf{Z})$ has an approximately linear drift:
$$d\tilde{\mathbf{X}} \approx \mathbf A \tilde{\mathbf{X}}dt +b(\tilde{\mathbf{X}})d\mathbf{W}$$
By applying the regression theorem (\ref{eq:RegresionTh}) the time dependent expectation values follows from which the relation (\ref{eq:virtualrespC}) can be obtained and used to finally derive the response functions of the true degrees of freedom.

%As a benchmark, we present an example where a larger list of observables is essential for accurately estimating the correct linear response and for which the calculation can be carried out analytically. Consider the following nonlinear stochastic differential equation:
\red{We begin with an instructive example in which incorporating a broader set of observables is crucial for accurately estimating the correct linear response. This example is particularly valuable as it allows for analytical calculations. Specifically, we consider the following nonlinear stochastic differential equation:
}
\begin{equation}\label{eq:toyexample}
\begin{cases}
    d \mathbf X = \mathbf G_{x} \mathbf X dt + \sigma_x d\mathbf{W}_x \\
    d \mathbf Y = \mathbf G_{y} \mathbf Y dt + \mathbf G_{yx} \mathbf X \odot \mathbf X + \sigma_y d\mathbf{W}_y
\end{cases}
\end{equation}
where $\odot$ indicates the element-wise product. We have two coupled systems, where the term $\mathbf{G}_{yx} \mathbf{X} \odot \mathbf{X}$ introduces nonlinear interactions between the variables. When considering the case in which $\mathbf{G}_{x}$ is a diagonal matrix, this system can be transformed into a linear one by introducing a new set of observables defined as $\mathbf{Z} = \mathbf{X} \odot \mathbf{X}$, leading to the following evolution
\begin{equation}\label{eq:ExtendedSDE}
\begin{cases}
    d \mathbf Z = 2 \mathbf G_{x} \mathbf Z dt + \sigma_x^2 dt + 2 \sigma_x \mathbf{X} \odot d\mathbf{W}_x \\
    d \mathbf X = \mathbf G_{x} \mathbf X dt + \sigma_x d\mathbf{W}_x \\
    d \mathbf Y = \mathbf G_{y} \mathbf Y dt + \mathbf G_{yx} \mathbf Z + \sigma_y d\mathbf{W}_y
\end{cases}
\end{equation}

Thanks to the regression theorem (\ref{eq:RegresionTh}) these systems admit an exact  finite-size Koopman matrix i.e. $\mathbf K (t)= e^{\mathbf A t}$, where $\mathbf{A}$ is the matrix of the linear drift in (\ref{eq:ExtendedSDE}). Then using \eqref{eq:CentralResult}, we can derive analytically the exact response function of \eqref{eq:toyexample}.
We analyze the agreement of equation \eqref{eq:MainResult} with the exact linear response curves for two sets of observables: i) Hermite functions up to order 1, referred to as $\mathrm{H}_1$; ii) Hermite functions up to order 2, referred to as $\mathrm{H}_2$ and compare both also to the response function sampled by perturbing the initial conditions, i.e. using the definition \eqref{eq:responseDef} and referred to as sampled response. In Fig. \ref{fig:1}a an example comparison between these three cases and the exact response function is showed.
The average error, see Fig. \ref{fig:1}b, associated with $\mathrm{H}_2$ scales with the number of realizations in the same way of the sampled response while $\mathrm{H}_1$ is order of magnitude worse. 
This happens because the stationary distribution $p_s(\bm x)$ is asymmetric thus a linear approximation, implicit in $\mathrm{H}_1$, will not work. However we note that the increase of the standard deviation $\sigma_y$ in \eqref{eq:stocSystem} mitigates the influence of $\mathrm X$ on $\mathrm Y$, thereby reducing the asymmetry of its distribution and leading the system to behave more linearly. This effect is clearly shown plotting the average error vs $\sigma_y$ (Fig. \ref{fig:1}c), with the accuracy of $\mathrm{H}_1$ approaches that of the sampled response as $\sigma_y$ increases.
\begin{figure}[ht!]
    \centering
    \includegraphics[width=0.47\textwidth]{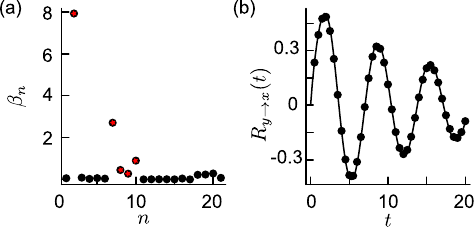}
    \caption{\red{(a) Weight coefficients $\beta_n$ for the Stuart-Landau system \eqref{eq:Stuart-Landau}. Dominant terms (highlighted in red) were selected as basis functions for the response function estimation. (b) Comparison between the estimated response function (derived from the reduced basis) and the exact analytical result.}}
    \label{fig:beta1}
\end{figure}

\begin{figure}[ht!]
    \centering
    \includegraphics[width=0.47\textwidth]{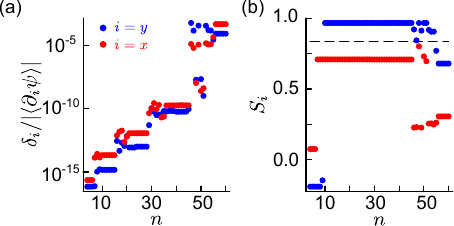}
    \caption{\red{(a) Normalized residual error for variables $x$ and $y$ in the gradient system \eqref{eq:GradSystem}, plotted against dictionary size. (b) Estimated entropy from eq. \eqref{eq:ConstrainEntropy} as a function of dictionary size. The dashed black line denotes the exact theoretical entropy computed from the stationary distribution.}}
    \label{fig:beta2}
\end{figure}
From a computational standpoint, equation \eqref{eq:MainResult} has been evaluated using a pseudoinverse operation for enhanced efficiency. Specifically, let $\mathbf V(t)$ represent the matrix whose columns correspond to the system variables at time $t$, with rows reflecting their values across different realizations of noise, and similarly let $\mathbf P(t)$ be the corresponding matrix for the set of functions $\psi$, the virtual response can be estimated as follow
\begin{equation*}
    \mathbf{\tilde{R}}_{\psi \rightarrow x}(t) = \left [ \mathbf V (t) \mathbf P^T (0) \right ] \left [ \mathbf P(0) \mathbf P ^{T}(0) \right ]^{-1} = \mathbf V (t) \mathbf P ^{\dagger} (0)
\end{equation*}
where $\{\cdot \}^{\dagger}$ denotes the pseudoinverse operation.
\red{For the example we have used so far, a basis of functions containing polynomials up to the second order is sufficient to achieve perfect agreement between equation (\ref{eq:MainResult}) and the exact response function. Clearly, this is not true in general. We selected three prototypical stochastic systems in two dimensions $(x,y)$ and computed the response of variable $x$ to a perturbation of $y$ i.e. $R_{y \to x}(t)$. We considered a gradient system with four stable states \cite{klus2018kernel}:
\begin{equation}\label{eq:GradSystem}
\begin{split}
        dX&=-\partial_XV(X,Y) dt +\sigma dW_x\\
    dY&=-\partial_YV(X,Y) dt +\sigma dW_y
\end{split}
\end{equation}
with $V(x,y)=(x-1)^2 +(y-0.5)^2 + 0.1 xy$. For a case with non polynomial drift, a network of two interacting neural population:
\begin{equation}\label{eq:TanhSystem}
\begin{split}
        dX&=[\tanh(J_{11}X +J_{12}Y)-X]dt +\sigma dW_x\\
    dY&=[\tanh(J_{21}X +J_{22}Y)-Y]dt +\sigma dW_y
\end{split}
\end{equation}
and finally a Stuart-Landau stochastic oscillator:
\begin{equation}\label{eq:Stuart-Landau}
\begin{split}
        dX&=[-a \left(X^2+Y^2\right) (b Y+X)+a X-Y]dt +\sigma dW_x\\
    dY&=[-a \left(X^2+Y^2\right) (b X+Y)+a Y+X]dt +\sigma dW_y
\end{split}
\end{equation}
These three systems presents qualitatively different stationary distribution $P(x,y)$, see Fig. \ref{fig:2DCases}, however that for all of them including in the function library  $\mathbf{\psi}(x,y)$ polynomial of total order 2 ($H_2$) had almost no improvements with respect to the linear case ($H_1$). At the same time however, including polynomial of order 3 ($H_3$) allows to fully capture the response function for all cases. The reason behind this behavior can be promptly understood for systems like (\ref{eq:GradSystem}). In fact for gradient systems the generalized F.D. theorem (\ref{eq:GeneralizedFD}) gives us:
$$R_{y\to x}(t) =2\langle X(t) \partial_Y V(X_0,Y_0) \rangle/\sigma^2\, . $$
In our case: $\partial_y V(x,y)=\alpha x+ \left(y^2-1/2\right) y$ which shows that the response is fully predicted once we consider the simple basis $\left \{ x,y,y^3 \right \}$.
However finding a good basis is generally a non trivial problem. In the following section we will address the problem of selecting the dictionary $\psi_n(\mathbf{x})$ in a principled way for general stochastic systems.
\section*{Principled selection of the dictionary of basis functions} 
We will occasionally use Einstein notation for the sums involved.
Equation (\ref{eq:MainResult}) suggest us a way to estimate the optimal base from empirical data. For all the two dimensional systems considered above the response of a variable $x$ due to a perturbation of $y$ is approximated by:
$$R_{y \to x}(t)\approx  \langle X \psi_n(\vec{X}_0) \rangle \beta_{n,2}$$
We defined:
$$\beta_{n,j} =[\mathbf{C}_{\mathbf{\psi,\psi}}(0)^{-1}]_{nk}\langle \partial_j\psi_{k} (\vec{X_0}) \rangle$$
where $(\partial_x,\partial_y)=(\partial_1,\partial_2)$.
All the integral involved in the weights $\beta$ are with respect to the stationary distribution thus can be generally computed empirically. As in all spectral approaches $\beta$ tells us the most relevant ``modes" in the expansion thus it allows us to optimally tune the dictionary of basis function $\mathbf{\psi}(\vec{x})$. A concrete example is shown in Fig. \ref{fig:beta1} where we computed $\beta_{n,2}$ with the usual basis of Hermite polynomials for the stationary state of Stuart-Landau system (\ref{eq:Stuart-Landau}). As shown in Fig. \ref{fig:beta1}a the only functions with a non negligible $\beta$ are 
\begin{equation*}
\begin{split}
    \mathbf{\psi}(x,y)=\{&H_1(x),H_1(y),H_1(x)H_2(y),\\ &H_2(x)H_1(y),H_3(x),H_3(y) \}
\end{split}
\end{equation*}
As excepted using this reduced base we get an excellent agreement in the response function, see Fig. \ref{fig:beta1}b.
\begin{figure*}[ht!]
    \centering
    \includegraphics[width=\textwidth]{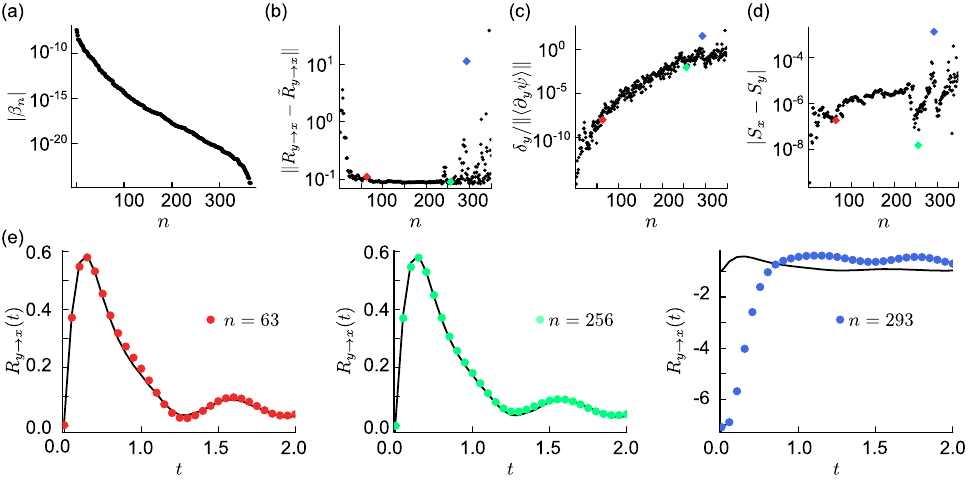}
    \caption{\red{(a) Sorted absolute values of coefficients  $|\beta_n|$ for the stochastic Lorenz system. (b) Relative error between exact response and $n$-element dictionary estimates. (c) Residual error response approximations. (d) Difference between estimated entropy values as in eq. (\ref{eq:dS0}). (e) Comparison of exact response (black curve) and approximated responses (colored scatter points) across three dictionary sizes.}}
    \label{fig:Lorenz}
\end{figure*}
The optimal choice of functions in general is the one that diagonalize the matrix $\mathbf{C}_{\psi,\psi}(0)$ or in other terms we should find a set of functions which is orthogonal with respect to the stationary measure $d\mu=P(\mathbf{x})d\mathbf{x}$. This is an old idea that dates back to the theory of polynomial chaos first introduced by Wiener and subsequent formal results such as the Kosambi–Karhunen–Loève theorem \cite{ghanem2003stochastic}. Even though the core concepts remained the same, these spectral expansion approaches, under the large umbrella of the Koopman formalism, have found recently renovated interest thanks to the widespread computational power and the development of new powerful machine-learning algorithm \cite{klus2018kernel}. We note however that typically in all these cases the focus is on learning a large set of approximated eigenfunction and eigenvalues of the Koopman operator from empirical measurements, which is undoubtedly challenging especially for high dimensional systems. In this work however we are interested only to specific response functions. For example we have seen that for system (\ref{eq:Stuart-Landau}) a basis of just 5 elements predicts perfectly $R_{y \to x}(t)$ but clearly, by no means this small dictionary is sufficient to represent a large collection of the true eigenfunctions of the Koopman operator. So far we have avoided questions related to optimal size of the dictionary of $\psi(\mathbf{x})$. A first reasonable step that allows to find a sparse dictionary is to compute $\beta$ from a large set of functions and sort the most relevant elements  with respect to the entries of $|\beta_n|$. We need however some guiding principle to properly truncate this sorted dictionary. It is possible to derive several analytical constrains that can guide this selection and optimization of the dictionary. In fact by comparing our expression (\ref{eq:MainResult}) with the F.D. theorem (\ref{eq:GeneralizedFD}) we find the central consistency relation:
\begin{equation}\label{eq:Consistency}
 \mathbf{\beta}\psi(\mathbf{x}) \approx -\mathbf{\nabla} \log(P(\mathbf{x}))
\end{equation}
By multiplying by an arbitrary function $f(\mathbf{x})$ and integrating with respect to the invariant measure we arrive at empirically testable relation:
\begin{equation}\label{eq:ConstrainF}
    \beta\langle \mathbf{\psi}(\mathbf{x}) f(\mathbf{x})\rangle \approx \langle \nabla f(\mathbf{x}) \rangle
\end{equation}
that can be used as an error estimation for the optimal choice of the basis. The arbitrary function $f(\mathbf{x})$ can be tailored to specific problem at hand. Notably, for the specific case $f(\mathbf{x})=\psi(\mathbf{x})$,  eq. (\ref{eq:ConstrainF}) translates to:
$$C_{\psi \psi}(0)\beta \approx \langle \nabla\psi \rangle \, .$$
The error:
\begin{equation}\label{eq:RegressError}
    \delta_i = \| C_{\psi,\psi}(0)\beta_i - \langle \partial_i \psi \rangle \|
\end{equation}
is minimized by the linear regression $\beta =C_{\psi \psi}(0)^{-1} \langle \nabla\psi \rangle$ which, consistently is also the definition of $\beta$ based on eq.(\ref{eq:MainResult}).\\
We can find a further constrains introducing the primitives of the dictionary with respect to the $j$-th variable $\Psi_{n,j}(\mathbf{x})$ i.e.:
$$\frac{d\Psi_{n,j}(\mathbf{x})}{dx_j}=\psi_n(\mathbf{x})$$
Finally obtaining the following approximation from (\ref{eq:Consistency}):
\begin{equation}\label{eq:ConstrainEntropy}
    \sum_{n} \beta_{n,j}\langle\Psi_{n,j}(\mathbf{x}) \rangle \approx -\int P(\mathbf{x})log[P(\mathbf{x})] d \mathbf{x} \, .
\end{equation}
The right hand side is the Shannon entropy $S$ of the stationary measure. Thus the proper choice of basis function must minimize also the error on the entropy:
$$S_j =\sum_{n} \beta_{n,j}\langle\Psi_{n,j}(\mathbf{x}) \rangle \approx S$$
Albeit estimating the stationary entropy from empirical measures is in general a challenging problem \cite{Vulpiani2009} several numerical approaches have been proposed in the years \cite{namdari2019review}. As a proof of concept we will use the gradient system (\ref{eq:GradSystem}) for which the stationary distribution, hence the entropy $S$, is know analytically. As we increase the size of the dictionary  we see in Fig. \ref{fig:beta2}a that the residual error increases, due to the increasing numerical error in the correlations matrix inversion. For this reason is  difficult to use this constrain alone as stop criteria for the convergence of the response. On the contrary the estimated entropy for both $x$ and $y$ quickly saturates close to the exact value when the cubic terms are introduced in the basis dictionary in accordance with the observation we made in the previous section. Even if we do not have in general access to the true value of $S$, the constrains (\ref{eq:ConstrainEntropy}) apply for all degrees of freedom thus it must hold in general that
\begin{equation}
    S_i -S_j \approx0 \, .
    \label{eq:dS0}
\end{equation}
}
\red{We tested this idea for the challenging case of the Lorenz system with additional uncorrelated white noise, as in \cite{baldovin2021handy}:}
\red{
\begin{equation}\label{eq:LorenzSystem}
\begin{split}
        dX &= [\sigma(Y-X)]dt +\sqrt{2\epsilon}dW_x \\
        dY &= [X(\rho -Z) - Y]dt +\sqrt{2\epsilon}dW_y \\
        dZ &= [XY-\beta Z]dt  +\sqrt{2\epsilon}dW_z 
\end{split}
\end{equation}
with the parameters of the standard chaotic case ($\sigma,\rho,\beta$= $10,28, 8/3$) and noise $\epsilon=8.0$ \footnote[2]{We verified that our results holds also for the case of small noise $\epsilon \sim 0.1$ but convergence requires almost double the number of dictionary functions, likely due the singular nature of the stationary measure}. With the aim of reproducing the response function $R_{y\to x}(t)$ we sampled $\mathbf{C_{\psi \psi}(0)}$ and $\langle \nabla\psi \rangle$ using a large basis of function with polynomials up to order 11. Sorting the resulting $\beta$ by absolute value we observe in Fig.\ref{fig:Lorenz}a we see a sort of continuum of $\beta$ witch makes difficult to select an a priori threshold for the dictionary size. The distance from the exact response function, obtained by perturbing the dynamics, and the one estimated using our approach quickly drops as more and more elements are included in the dictionary, see Fig. \ref{fig:Lorenz}b. At the same time however we see that after a plateau adding more elements in the dictionary results in larger errors in the response estimate. Also in this case the residual error alone, Fig. \ref{fig:Lorenz}c, it is not enough to decide when to stop because it increases on average with the dictionary size. On the other hand, the difference in entropy estimations gives very clear signals, see Fig. \ref{fig:Lorenz}d especially when the elements of the dictionary are too many. To show this we selected three different sizes $n=63$ (red), $n=256$ (green) and $n=293$ (blue) which correspond respectively to two local minima and a local maxima of $|S_x -S_y|$. Convergence in the response is already found at $n=63$, see Fig. \ref{fig:Lorenz}e, as correctly signaled. This also applies for the much larger basis of $n=256$. On the contrary for the case $n=293$, where the entropy difference is maximal, also the error in the response function estimate is large.}

\red{All our observations, including the self-consistent constrains that can guide the dictionary optimization, open the door to an effective way to empirically estimate causality between different degrees of freedom from empirical measurement of unperturbed stochastic systems as we will see in the next section.}

\section*{From response to causality }
\begin{figure*}[ht!]
    \centering
    \includegraphics[width=180mm]{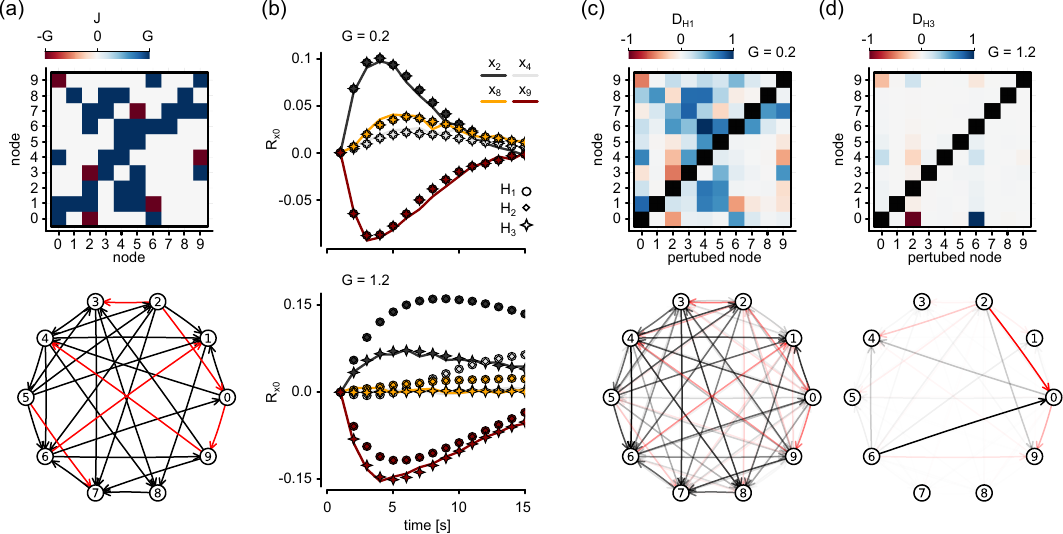}
    \caption{\red{(a), Top: Synaptic matrix for the ten-dimensional example system described in eq. (\ref{eq:reservoir}). Here, the parameter G scales the connection strengths, thereby modulating the system’s nonlinearity. Bottom: The corresponding directed graph, where red arrows indicate connections with negative weights.
    (b), Response curves following perturbations to the first node ($x_1$) for two different levels of nonlinearity. Solid lines show the estimates obtained via the perturbative approach, scatter points represent the estimates from \eqref{eq:MainResult} using observables from the Hermite basis $\mathrm H_1$, $\mathrm H_2$ and $\mathrm H_3$.
    (c-d), Estimated causality matrices (top panels) and their associated graphs (bottom panels) for parameter values $G = 0.2$ (c) and $G = 1.2$ (d). In the graphs, red edges denote negative causal dependencies, and the transparency of each edge reflects the intensity of the causal connection.
    The system is simulated for $50 \, s$, the time resolution is set at $\mathrm{dt}=0.01 \, s$, perturbation amplitude $\epsilon=0.01$, with stochastic Heun method for integration.}
    }
    \label{fig:4}
\end{figure*}

\begin{figure}
    \centering
    \includegraphics[width=0.5\textwidth]{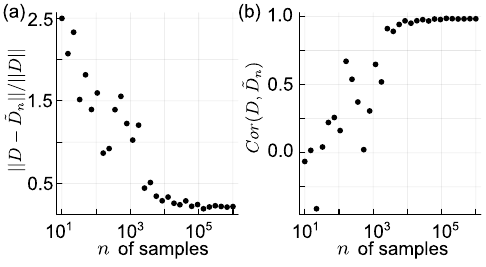}
    \caption{\red{Relative distance measured by the Frobenius norm of the real matrix $D$ and the one estimated from $n$ samples $\tilde{D}_n$ is shown in (a) while element wise correlation is reported in (b).}}
    \label{fig:Samples}
\end{figure}

It is possible to use the response function as a measure of interventional causality. Following \cite{baldovin2020understanding}, we can measure the strength of the causal dependence of the variable $x_{i}$ on the variable $x_j$ as the total response that the former has consequently a perturbation of the latter i.e.:
\begin{equation}\label{eq:DefD}
    \mathrm D_{ij}=\int_{0}^{\infty} \mathrm R_{x_j \to x_i}(t) dt
\end{equation}
then using our result \eqref{eq:MainResult} we have:
\begin{equation}
   \mathbf{D} \approx \left[ \int_{0}^{\infty} \mathbf{C_{x\psi}}(t) dt \right ] \mathbf{C_{\psi \psi}}(0)^{-1} \braket{\nabla \bm \psi}
\end{equation}
Thus relating correlation and causality. The better is the estimation of the response function the better the estimation of the causal link between variables will be. 
It is also noteworthy that by selecting the eigenfunctions $\phi$ of $\mathcal{L}$ as the basis, we derive: 

$$ \mathrm D_{ij} \approx - \sum_{k} \alpha_{ik} \frac{\braket{ \partial_{j}\phi_k}}{\lambda_k}$$

where $\lambda_k$ denotes the eigenvalues of $\mathcal{L}$. In essence, this indicates that the causal relationship $\mathrm D_{ij}$, can be interpreted as a summation of contributions from $x_j$ across the all eigenmodes. These are quantified as $\braket{ \partial_{j}\phi_k (\bm k)}$ and are weighted by the significance of each eigenmode, i.e. the associated timescale $1/\lambda_k$. 
\red{To investigate the behavior of this causality measure in complex high-dimensional systems, we consider the prototype of a biologically inspired network of neurons \cite{Treves1993, Mattia2002}. The system's state is represented by the vector $\mathbf{X}(t)$, whose element $\mathrm{X}_i(t)$ corresponds to the activity of the $i$-th unit, evolving according to the nonlinear dynamics
\begin{equation}
    d \mathbf{X} = [\tanh(G\mathbf{J} \mathbf{X}) - \mathbf{X}]dt + \sigma d\bm{W}
    \label{eq:reservoir}
\end{equation}
where $\mathbf{J} \in \mathbb{R}^{N \times N}$ is the synaptic matrix governing the recurrent input currents, and $\bm{W}(t)$ is $N-$dimensional uncorrelated Wiener process. This dynamical framework also serves as the foundation for several modern machine learning approaches \cite{Jaeger2004, pathak2018model, di2024linearizing}.
For our test set, we consider $N=10$ nodes, sparsely connected with a synaptic strength $\mathrm{G}$ (Fig. \ref{fig:4}a). These connections may be excitatory (positive) or inhibitory (negative), influencing the dynamics. The factor $\mathrm{G} \in \mathbb{R}$ modulates the recurrent current, thereby controlling the level of nonlinearity in the system.
Although the connection pattern remains invariant, the system's response can vary significantly based on the value of $\mathrm{G}$. As expected, for small values of $\mathrm{G}$, the system behaves nearly linearly, and the first-order observables $H_1$ can adequately capture the linear response. For larger values of $\mathrm{G}$, the third-order term of the hyperbolic tangent function becomes crucial for describing the system's behavior, necessitating the inclusion of higher-order terms, as shown in Fig. \ref{fig:4}b.
In cases with low nonlinearity, the causality links closely resemble the synaptic matrix (Fig. \ref{fig:4}c). However, as nonlinearity increases, the information in the connectivity matrix diminishes. As seen in Fig. \ref{fig:4}d, for high values of $\mathrm{G}$, the primary causal links reduce in number, further amplified by the nonlinearity. As a result, paths of higher lengths become crucial and a node without a direct connection may exert a stronger influence than nodes with direct connections. For example, in Fig. \ref{fig:4} d, we observe an inhibitory influence of node ``6" on node ``9", despite the positive connection, and a new excitatory influence emerges on node ``4". All our results are implicitly based on the capabilities of estimating expectation values from a large collection of samples. In practical context however the number of samples is usually limited, thus it's important to verify the convergence of the causality matrix as the sample size is increased. In Fig. (\ref{fig:Samples}) we show that for system (\ref{eq:reservoir}) already $10^4$ samples are enough to converge to the right $\mathbf{D}$. While the rate of convergence depends strongly on the specific systems under exam, this shows that even for such an high dimensional systems, a small sample size might be enough to infer proper causal relations.}

\section*{Conclusions  and future directions }
By establishing a link between the response function and the Koopman operator we derived an alternative way to compute the response function for a general class of non-linear stochastic systems. Our result is rooted in spectral approach that, albeit quite old \cite{koopman1931hamiltonian,neumann1932operatorenmethode}, is now facing renewed interest and widespread adoption \cite{brunton2022modern,budivsic2012applied}. Besides all the possible issues that are common to causality inference as a general, such as non-markovianity \cite{capone2018spontaneous} and non-stationarity, the major limitation of our result is in the proper choice of the library of basis functions. In fact while convergence can be formally guaranteed when the size of the library goes to infinity, in practice the best would be to find the smallest set of function possible to avoid spectral pollution and over-fitting. A small size in the library is also essential in systems of high dimensions to avoid computational cost. \red{We found several self consistent constrains that strongly guide dictionary selection and optimization as we showed for several examples.}  While we will leave to a future work \red{further} investigation of related issues we would like to point toward \red{interesting opportunities} offered by the great approximation capabilities of machine learning algorithm and deep-neural network. For example in \cite{lusch2018,alford2022deep,Otto,brunton2022modern} the authors showed that already a simple architecture like an auto-encoder is capable to find, by minimizing the loss function
$$|| \mathrm{E}[\bm \psi (\mathbf X_t) | \mathbf X_0] -\mathbf K(t) \bm \psi(\mathbf X_0) ||_2$$
or its variants, a satisfactory low-dimensional representation of the Koopman operator for the chaotic Lorenz dynamical system and other non-linear examples. \red{The constrains we found can equally be used as loss function in the same settings}. The advantage of using simple architectures is that, once the weights of the network are known, the matrix $ \partial _{x_i} \psi_j (\bm x)$ can be computed analytically. In general, it is true that the challenges inherent in applying the Koopman framework—or, more broadly, any pseudo-spectral decomposition—also affect our approach. However, the regression theorem and our examples demonstrate that it may be possible, in principle, to identify a small number of virtual degrees of freedom that approximately linearize the drift. This does not imply that we have fully reconstructed the entire spectrum of the Koopman operator, which is clearly infinite. Nonetheless, our results suggest that it might still be feasible to achieve highly accurate estimates of the response functions \red{and consequently of causal links}.
In other words, an incomplete reconstruction of the Koopman operator can still lead to an exact reproduction of the response functions for appropriately chosen degrees of freedom—a point that should be considered when dealing with high-dimensional systems \red{as we have showed here for several examples}.
\red{The causal measure explored in this paper is highly adaptable to various extensions. Depending on the time scale of interest, one can limit the response integral to different time ranges, looking for example for short-term and long-term causal links.
Furthermore, dynamical systems can exhibit different behaviors depending on the basin of attraction considered. By sampling trajectory data from specific regions of the state space, the analysis can be further tailored to particular conditions.}
Being able to infer the causal network of a set of measured variables without the need of perturbing the system is certainly powerful. Thus, given the general premises and the data-driven nature of our results
we expect that this approach will be a valuable tool and will find several application in multiple domain of science.
\begin{acknowledgments}
We are grateful to Maurizio Mattia, Guido Gigante, Roberto Benzi, Angelo Vulpiani and Marco Baldovin for their valuable suggestions and comments on an earlier version of this manuscript. Work partially funded by the Italian National Recovery and Resilience Plan (PNRR), M4C2, funded by the European Union - NextGenerationEU (Project IR0000011, CUP B51E22000150006, ‘EBRAINS-Italy’) to Maurizio Mattia.
\end{acknowledgments}

\appendix

\section{The Koopman matrix for system \eqref{eq:ExtendedSDE}}
Consider the general system with a linear drift and arbitrary diffusion and natural boundary conditions:
\begin{equation}
    d X_i = \bm A_{ij}  X_j dt + b_{ij}(\bm X) dW_j
\end{equation}
where we used Einstein notation for repeated indices and $dW_j$ are independent Wiener noise sources. Defined the diffusion matrix as $\bm D =\bm b ^T \bm b /2$, the evolution in time of probability density $p(\bm X,t | \bm X_0)$ is governed by the following Fokker-Planck equation \cite{risken1996fokker}:
\begin{equation}\label{eq:Fokker-PlanckLinearDrift}
    \partial_t p =- A_{ij} \partial _i \left ( x_j p \right) +\partial_i \partial j \left( D_{ij} p \right )
\end{equation}
multiplying \eqref{eq:Fokker-PlanckLinearDrift} by $x_k$ and taking the average $\braket{}$ and apply integration by parts.
The diffusion part is zero even in multiplicative case since:
$$\int x_k \partial_i \partial_j D_{ij}p =\int \partial_i \partial_j(x_k)D_{ij}p =0 $$
Thus that moments follows the linear dynamics:
\begin{equation}
    \partial_t \mathrm E[X_{i,t}|\mathbf{X}_0] = A_{ij}\mathrm E[X_{j,t}|\mathbf{X}_0]
\end{equation}
A result commonly refereed to as regression theorem \cite{gardiner2009stochastic}. The solution can be written using the exponential matrix: 
$$\mathrm E[X_{i,t}|\mathbf{X}_0] = e^{\bm A t} \mathbf{X}_0$$
thus following the definition, the Koopman matrix for this set of observable is: $ \bm K (t)= e^{\bm A t}$

\section{Response function from Koopman decomposition}
Consider a set of $d$ observables $\{\psi_k(\bm x)\}_{k=1}^d$ ideally spanning a subspace containing the state of the system at all times and that admits the family of $d \times d$ matrices $\mathbf K(t)$ that approximate the action of the Koopman operators $\mathcal{K}^t$.

The average observable evolution in this space, with initial condition $\mathbf X_0$ is described as
\begin{equation}
    \mathrm{E}[\bm \psi (\mathbf X_t | \mathbf X_0] \approx \mathbf K(t) \bm \psi (\mathbf X_0)
    \label{eq:Konpsi}
\end{equation}
Under the semi-group assumption it would hold the relation $\mathbf K(t) = e^{\mathbf L t}$ with $\mathbf L$ approximating the action of the infinitesimal Koopman generator $\mathcal{L}$. Note that for $d \to \infty$ convergence is guaranteed and $ \mathbf L$ becomes iso-spectral to $\mathcal{L}$.
In the particular case in which the observables $\psi_k$ are the linear evolving eigenfunctions $\phi_k$, $\mathbf K$ would be the diagonal matrix of the associated eigenvalues $\lambda_k$.

Given the coordinates of the state of the system in this space $\bm x = \bm \alpha \bm \psi$, the expected value of the associated stochastic process up to time t is the following

\begin{equation}
    \mathrm{E}[\mathrm X_{i, t} | \mathbf X_0] = \sum_{n=1}^d \sum_{m=1}^d  \alpha_{in} \mathrm K_{nm} (t) \psi_m (\mathbf X_0)
\end{equation}

The effect on the physical variable $x_j$ of perturbing the function (virtual variable) $\psi_i[\mathbf X_0] \rightarrow \psi_i[\mathbf X_0] + \epsilon_{\psi}$ is then described as
\begin{equation*}
    \mathrm{E}_p[\mathrm X_{j, t} | \mathbf X_0] = \,  \mathrm{E}[\mathrm X_{j, t} | \mathbf X_0] + \tilde{\mathrm R}_{\psi_i \rightarrow x_j}(t) \, \epsilon_{\psi}
\end{equation*}
where we introduce $\tilde{\mathrm R}_{\psi_i \rightarrow x_j}(t)$ as the virtual Response function defined by
\begin{equation*}
    \tilde{\mathrm R}_{\psi_i \rightarrow x_j}(t) = \lim_{\epsilon_{\psi} \to 0} \frac{\langle  \mathrm X_j(t)\rangle _p -\langle  \mathrm X_j(t)\rangle}{\epsilon_{\psi}} = [\bm \alpha \mathbf K (t)]_{ji}
\end{equation*}

where the brackets include the average with respect to the distribution of the initial conditions.
Note that for $\psi_k$ eigenfunctions and semi-group condition, it holds $\tilde{\mathrm R}_{\psi_i \rightarrow x_j}(t) = \alpha_{ji}e^{\lambda_i t} $.

Let's now look at physical perturbations. Let's consider the perturbation $\mathrm X_{i, 0} \rightarrow \mathrm X_{i, 0} + \epsilon$. Expanding $\psi_k[\mathbf X_t]$ around the unperturbed system we find
\begin{equation*}
    \mathrm{E}_p[\mathrm X_{j,t} | \, \mathbf X_0] = 
    \mathrm{E}[\mathrm X_{j, t} | \, \mathbf X_0] + \epsilon \sum_{m=1}^d  \left.  \frac{\partial \psi_m(\mathbf x)]}{\partial x_i} \right |_{\mathbf X_0} [\bm \alpha \mathbf K (t)]_{jm}
\end{equation*}
\begin{equation*}
    =\mathrm{E}[\mathrm X_{j, t} | \, \mathbf X_0] + \epsilon \sum_{m=1}^d \left. \frac{\partial \psi_m(\mathbf x)]}{\partial x_i} \right |_{\mathbf X_0} \tilde{\mathrm R}_{\psi_m \rightarrow x_j}(t)
\end{equation*}

Averaging over the initial condition distribution, by the linear response definition we find a relation that links the physical to the virtual response function
\begin{equation}
    \text{R}_{x_i\rightarrow x_j}(t) = \sum_{m=1}^d \left \langle \left. \frac{\partial \psi_m(\mathbf x)}{\partial x_i} \right |_{\mathbf X_0} \right \rangle \tilde{\mathrm R}_{\psi_m \rightarrow x_j}(t)
\end{equation}

\bibliography{RefLibrary}% Produces the bibliography via BibTeX.

\begin{thebibliography}{43}%
\makeatletter
\providecommand \@ifxundefined [1]{%
 \@ifx{#1\undefined}
}%
\providecommand \@ifnum [1]{%
 \ifnum #1\expandafter \@firstoftwo
 \else \expandafter \@secondoftwo
 \fi
}%
\providecommand \@ifx [1]{%
 \ifx #1\expandafter \@firstoftwo
 \else \expandafter \@secondoftwo
 \fi
}%
\providecommand \natexlab [1]{#1}%
\providecommand \enquote  [1]{``#1''}%
\providecommand \bibnamefont  [1]{#1}%
\providecommand \bibfnamefont [1]{#1}%
\providecommand \citenamefont [1]{#1}%
\providecommand \href@noop [0]{\@secondoftwo}%
\providecommand \href [0]{\begingroup \@sanitize@url \@href}%
\providecommand \@href[1]{\@@startlink{#1}\@@href}%
\providecommand \@@href[1]{\endgroup#1\@@endlink}%
\providecommand \@sanitize@url [0]{\catcode `\\12\catcode `\$12\catcode
  `\&12\catcode `\#12\catcode `\^12\catcode `\_12\catcode `\%12\relax}%
\providecommand \@@startlink[1]{}%
\providecommand \@@endlink[0]{}%
\providecommand \url  [0]{\begingroup\@sanitize@url \@url }%
\providecommand \@url [1]{\endgroup\@href {#1}{\urlprefix }}%
\providecommand \urlprefix  [0]{URL }%
\providecommand \Eprint [0]{\href }%
\providecommand \doibase [0]{https://doi.org/}%
\providecommand \selectlanguage [0]{\@gobble}%
\providecommand \bibinfo  [0]{\@secondoftwo}%
\providecommand \bibfield  [0]{\@secondoftwo}%
\providecommand \translation [1]{[#1]}%
\providecommand \BibitemOpen [0]{}%
\providecommand \bibitemStop [0]{}%
\providecommand \bibitemNoStop [0]{.\EOS\space}%
\providecommand \EOS [0]{\spacefactor3000\relax}%
\providecommand \BibitemShut  [1]{\csname bibitem#1\endcsname}%
\let\auto@bib@innerbib\@empty
%</preamble>
\bibitem [{\citenamefont {Wiener}(1956)}]{wiener1956theory}%
  \BibitemOpen
  \bibfield  {author} {\bibinfo {author} {\bibfnamefont {N.}~\bibnamefont
  {Wiener}},\ }\bibfield  {title} {\bibinfo {title} {The theory of
  prediction},\ }\href@noop {} {\bibfield  {journal} {\bibinfo  {journal}
  {Modern mathematics for engineers}\ } (\bibinfo {year} {1956})}\BibitemShut
  {NoStop}%
\bibitem [{\citenamefont {Granger}(1969)}]{granger1969investigating}%
  \BibitemOpen
  \bibfield  {author} {\bibinfo {author} {\bibfnamefont {C.~W.}\ \bibnamefont
  {Granger}},\ }\bibfield  {title} {\bibinfo {title} {Investigating causal
  relations by econometric models and cross-spectral methods},\ }\href@noop {}
  {\bibfield  {journal} {\bibinfo  {journal} {Econometrica: journal of the
  Econometric Society}\ ,\ \bibinfo {pages} {424}} (\bibinfo {year}
  {1969})}\BibitemShut {NoStop}%
\bibitem [{\citenamefont {Kantz}\ and\ \citenamefont
  {Schreiber}(2003)}]{kantz2003nonlinear}%
  \BibitemOpen
  \bibfield  {author} {\bibinfo {author} {\bibfnamefont {H.}~\bibnamefont
  {Kantz}}\ and\ \bibinfo {author} {\bibfnamefont {T.}~\bibnamefont
  {Schreiber}},\ }\href@noop {} {\emph {\bibinfo {title} {Nonlinear time series
  analysis}}}\ (\bibinfo  {publisher} {Cambridge university press},\ \bibinfo
  {year} {2003})\BibitemShut {NoStop}%
\bibitem [{\citenamefont {Runge}\ \emph {et~al.}(2023)\citenamefont {Runge},
  \citenamefont {Gerhardus}, \citenamefont {Varando}, \citenamefont {Eyring},\
  and\ \citenamefont {Camps-Valls}}]{runge2023causal}%
  \BibitemOpen
  \bibfield  {author} {\bibinfo {author} {\bibfnamefont {J.}~\bibnamefont
  {Runge}}, \bibinfo {author} {\bibfnamefont {A.}~\bibnamefont {Gerhardus}},
  \bibinfo {author} {\bibfnamefont {G.}~\bibnamefont {Varando}}, \bibinfo
  {author} {\bibfnamefont {V.}~\bibnamefont {Eyring}},\ and\ \bibinfo {author}
  {\bibfnamefont {G.}~\bibnamefont {Camps-Valls}},\ }\bibfield  {title}
  {\bibinfo {title} {Causal inference for time series},\ }\href@noop {}
  {\bibfield  {journal} {\bibinfo  {journal} {Nature Reviews Earth \&
  Environment}\ }\textbf {\bibinfo {volume} {4}},\ \bibinfo {pages} {487}
  (\bibinfo {year} {2023})}\BibitemShut {NoStop}%
\bibitem [{\citenamefont {Seth}\ \emph {et~al.}(2015)\citenamefont {Seth},
  \citenamefont {Barrett},\ and\ \citenamefont {Barnett}}]{setH2015granger}%
  \BibitemOpen
  \bibfield  {author} {\bibinfo {author} {\bibfnamefont {A.~K.}\ \bibnamefont
  {Seth}}, \bibinfo {author} {\bibfnamefont {A.~B.}\ \bibnamefont {Barrett}},\
  and\ \bibinfo {author} {\bibfnamefont {L.}~\bibnamefont {Barnett}},\
  }\bibfield  {title} {\bibinfo {title} {Granger causality analysis in
  neuroscience and neuroimaging},\ }\href@noop {} {\bibfield  {journal}
  {\bibinfo  {journal} {Journal of Neuroscience}\ }\textbf {\bibinfo {volume}
  {35}},\ \bibinfo {pages} {3293} (\bibinfo {year} {2015})}\BibitemShut
  {NoStop}%
\bibitem [{\citenamefont {Vinci}\ and\ \citenamefont
  {Benzi}(2018)}]{vinci2018economic}%
  \BibitemOpen
  \bibfield  {author} {\bibinfo {author} {\bibfnamefont {G.~V.}\ \bibnamefont
  {Vinci}}\ and\ \bibinfo {author} {\bibfnamefont {R.}~\bibnamefont {Benzi}},\
  }\bibfield  {title} {\bibinfo {title} {Economic complexity: Correlations
  between gross domestic product and fitness},\ }\href@noop {} {\bibfield
  {journal} {\bibinfo  {journal} {Entropy}\ }\textbf {\bibinfo {volume} {20}},\
  \bibinfo {pages} {766} (\bibinfo {year} {2018})}\BibitemShut {NoStop}%
\bibitem [{\citenamefont {Javarone}\ \emph {et~al.}(2023)\citenamefont
  {Javarone}, \citenamefont {Di~Antonio}, \citenamefont {Vinci}, \citenamefont
  {Cristodaro}, \citenamefont {Tessone},\ and\ \citenamefont
  {Pietronero}}]{javarone2023disorder}%
  \BibitemOpen
  \bibfield  {author} {\bibinfo {author} {\bibfnamefont {M.~A.}\ \bibnamefont
  {Javarone}}, \bibinfo {author} {\bibfnamefont {G.}~\bibnamefont
  {Di~Antonio}}, \bibinfo {author} {\bibfnamefont {G.~V.}\ \bibnamefont
  {Vinci}}, \bibinfo {author} {\bibfnamefont {R.}~\bibnamefont {Cristodaro}},
  \bibinfo {author} {\bibfnamefont {C.~J.}\ \bibnamefont {Tessone}},\ and\
  \bibinfo {author} {\bibfnamefont {L.}~\bibnamefont {Pietronero}},\ }\bibfield
   {title} {\bibinfo {title} {Disorder unleashes panic in bitcoin dynamics},\
  }\href@noop {} {\bibfield  {journal} {\bibinfo  {journal} {Journal of
  Physics: Complexity}\ }\textbf {\bibinfo {volume} {4}},\ \bibinfo {pages}
  {045002} (\bibinfo {year} {2023})}\BibitemShut {NoStop}%
\bibitem [{\citenamefont {Marconi}\ \emph {et~al.}(2008)\citenamefont
  {Marconi}, \citenamefont {Puglisi}, \citenamefont {Rondoni},\ and\
  \citenamefont {Vulpiani}}]{marconi2008fluctuation}%
  \BibitemOpen
  \bibfield  {author} {\bibinfo {author} {\bibfnamefont {U.~M.~B.}\
  \bibnamefont {Marconi}}, \bibinfo {author} {\bibfnamefont {A.}~\bibnamefont
  {Puglisi}}, \bibinfo {author} {\bibfnamefont {L.}~\bibnamefont {Rondoni}},\
  and\ \bibinfo {author} {\bibfnamefont {A.}~\bibnamefont {Vulpiani}},\
  }\bibfield  {title} {\bibinfo {title} {Fluctuation--dissipation: response
  theory in statistical physics},\ }\href@noop {} {\bibfield  {journal}
  {\bibinfo  {journal} {Physics reports}\ }\textbf {\bibinfo {volume} {461}},\
  \bibinfo {pages} {111} (\bibinfo {year} {2008})}\BibitemShut {NoStop}%
\bibitem [{\citenamefont {Baldovin}\ \emph {et~al.}(2020)\citenamefont
  {Baldovin}, \citenamefont {Cecconi},\ and\ \citenamefont
  {Vulpiani}}]{baldovin2020understanding}%
  \BibitemOpen
  \bibfield  {author} {\bibinfo {author} {\bibfnamefont {M.}~\bibnamefont
  {Baldovin}}, \bibinfo {author} {\bibfnamefont {F.}~\bibnamefont {Cecconi}},\
  and\ \bibinfo {author} {\bibfnamefont {A.}~\bibnamefont {Vulpiani}},\
  }\bibfield  {title} {\bibinfo {title} {Understanding causation via
  correlations and linear response theory},\ }\href@noop {} {\bibfield
  {journal} {\bibinfo  {journal} {Physical Review Research}\ }\textbf {\bibinfo
  {volume} {2}},\ \bibinfo {pages} {043436} (\bibinfo {year}
  {2020})}\BibitemShut {NoStop}%
\bibitem [{\citenamefont {Baldovin}\ \emph {et~al.}(2022)\citenamefont
  {Baldovin}, \citenamefont {Cecconi}, \citenamefont {Provenzale},\ and\
  \citenamefont {Vulpiani}}]{baldovin2022extracting}%
  \BibitemOpen
  \bibfield  {author} {\bibinfo {author} {\bibfnamefont {M.}~\bibnamefont
  {Baldovin}}, \bibinfo {author} {\bibfnamefont {F.}~\bibnamefont {Cecconi}},
  \bibinfo {author} {\bibfnamefont {A.}~\bibnamefont {Provenzale}},\ and\
  \bibinfo {author} {\bibfnamefont {A.}~\bibnamefont {Vulpiani}},\ }\bibfield
  {title} {\bibinfo {title} {Extracting causation from millennial-scale climate
  fluctuations in the last 800 kyr},\ }\href@noop {} {\bibfield  {journal}
  {\bibinfo  {journal} {Scientific Reports}\ }\textbf {\bibinfo {volume}
  {12}},\ \bibinfo {pages} {15320} (\bibinfo {year} {2022})}\BibitemShut
  {NoStop}%
\bibitem [{\citenamefont {Baldovin}\ \emph {et~al.}(2021)\citenamefont
  {Baldovin}, \citenamefont {Caprini},\ and\ \citenamefont
  {Vulpiani}}]{baldovin2021handy}%
  \BibitemOpen
  \bibfield  {author} {\bibinfo {author} {\bibfnamefont {M.}~\bibnamefont
  {Baldovin}}, \bibinfo {author} {\bibfnamefont {L.}~\bibnamefont {Caprini}},\
  and\ \bibinfo {author} {\bibfnamefont {A.}~\bibnamefont {Vulpiani}},\
  }\bibfield  {title} {\bibinfo {title} {Handy fluctuation-dissipation relation
  to approach generic noisy systems and chaotic dynamics},\ }\href@noop {}
  {\bibfield  {journal} {\bibinfo  {journal} {Physical Review E}\ }\textbf
  {\bibinfo {volume} {104}},\ \bibinfo {pages} {L032101} (\bibinfo {year}
  {2021})}\BibitemShut {NoStop}%
\bibitem [{\citenamefont {Lucente}\ \emph {et~al.}(2022)\citenamefont
  {Lucente}, \citenamefont {Baldassarri}, \citenamefont {Puglisi},
  \citenamefont {Vulpiani},\ and\ \citenamefont
  {Viale}}]{lucente2022inference}%
  \BibitemOpen
  \bibfield  {author} {\bibinfo {author} {\bibfnamefont {D.}~\bibnamefont
  {Lucente}}, \bibinfo {author} {\bibfnamefont {A.}~\bibnamefont
  {Baldassarri}}, \bibinfo {author} {\bibfnamefont {A.}~\bibnamefont
  {Puglisi}}, \bibinfo {author} {\bibfnamefont {A.}~\bibnamefont {Vulpiani}},\
  and\ \bibinfo {author} {\bibfnamefont {M.}~\bibnamefont {Viale}},\ }\bibfield
   {title} {\bibinfo {title} {Inference of time irreversibility from incomplete
  information: Linear systems and its pitfalls},\ }\href@noop {} {\bibfield
  {journal} {\bibinfo  {journal} {Physical Review Research}\ }\textbf {\bibinfo
  {volume} {4}},\ \bibinfo {pages} {043103} (\bibinfo {year}
  {2022})}\BibitemShut {NoStop}%
\bibitem [{\citenamefont {Rowley}\ \emph {et~al.}(2009)\citenamefont {Rowley},
  \citenamefont {Mezi{\'c}}, \citenamefont {Bagheri}, \citenamefont
  {Schlatter},\ and\ \citenamefont {Henningson}}]{rowley2009spectral}%
  \BibitemOpen
  \bibfield  {author} {\bibinfo {author} {\bibfnamefont {C.~W.}\ \bibnamefont
  {Rowley}}, \bibinfo {author} {\bibfnamefont {I.}~\bibnamefont {Mezi{\'c}}},
  \bibinfo {author} {\bibfnamefont {S.}~\bibnamefont {Bagheri}}, \bibinfo
  {author} {\bibfnamefont {P.}~\bibnamefont {Schlatter}},\ and\ \bibinfo
  {author} {\bibfnamefont {D.~S.}\ \bibnamefont {Henningson}},\ }\bibfield
  {title} {\bibinfo {title} {Spectral analysis of nonlinear flows},\
  }\href@noop {} {\bibfield  {journal} {\bibinfo  {journal} {Journal of fluid
  mechanics}\ }\textbf {\bibinfo {volume} {641}},\ \bibinfo {pages} {115}
  (\bibinfo {year} {2009})}\BibitemShut {NoStop}%
\bibitem [{\citenamefont {Klus}\ \emph {et~al.}(2020)\citenamefont {Klus},
  \citenamefont {N{\"u}ske}, \citenamefont {Peitz}, \citenamefont {Niemann},
  \citenamefont {Clementi},\ and\ \citenamefont {Sch{\"u}tte}}]{klus2020data}%
  \BibitemOpen
  \bibfield  {author} {\bibinfo {author} {\bibfnamefont {S.}~\bibnamefont
  {Klus}}, \bibinfo {author} {\bibfnamefont {F.}~\bibnamefont {N{\"u}ske}},
  \bibinfo {author} {\bibfnamefont {S.}~\bibnamefont {Peitz}}, \bibinfo
  {author} {\bibfnamefont {J.-H.}\ \bibnamefont {Niemann}}, \bibinfo {author}
  {\bibfnamefont {C.}~\bibnamefont {Clementi}},\ and\ \bibinfo {author}
  {\bibfnamefont {C.}~\bibnamefont {Sch{\"u}tte}},\ }\bibfield  {title}
  {\bibinfo {title} {Data-driven approximation of the koopman generator: Model
  reduction, system identification, and control},\ }\href@noop {} {\bibfield
  {journal} {\bibinfo  {journal} {Physica D: Nonlinear Phenomena}\ }\textbf
  {\bibinfo {volume} {406}},\ \bibinfo {pages} {132416} (\bibinfo {year}
  {2020})}\BibitemShut {NoStop}%
\bibitem [{\citenamefont {Zagli}\ \emph {et~al.}(2024)\citenamefont {Zagli},
  \citenamefont {Colbrook}, \citenamefont {Lucarini}, \citenamefont
  {Mezi{\'c}},\ and\ \citenamefont {Moroney}}]{zagli2024bridging}%
  \BibitemOpen
  \bibfield  {author} {\bibinfo {author} {\bibfnamefont {N.}~\bibnamefont
  {Zagli}}, \bibinfo {author} {\bibfnamefont {M.}~\bibnamefont {Colbrook}},
  \bibinfo {author} {\bibfnamefont {V.}~\bibnamefont {Lucarini}}, \bibinfo
  {author} {\bibfnamefont {I.}~\bibnamefont {Mezi{\'c}}},\ and\ \bibinfo
  {author} {\bibfnamefont {J.}~\bibnamefont {Moroney}},\ }\bibfield  {title}
  {\bibinfo {title} {Bridging the gap between koopmanism and response theory:
  Using natural variability to predict forced response},\ }\href@noop {}
  {\bibfield  {journal} {\bibinfo  {journal} {arXiv preprint arXiv:2410.01622}\
  } (\bibinfo {year} {2024})}\BibitemShut {NoStop}%
\bibitem [{\citenamefont {Gardiner}(2009)}]{gardiner2009stochastic}%
  \BibitemOpen
  \bibfield  {author} {\bibinfo {author} {\bibfnamefont {C.}~\bibnamefont
  {Gardiner}},\ }\href@noop {} {\emph {\bibinfo {title} {Stochastic
  methods}}},\ Vol.~\bibinfo {volume} {4}\ (\bibinfo  {publisher} {Springer
  Berlin Heidelberg},\ \bibinfo {year} {2009})\BibitemShut {NoStop}%
\bibitem [{\citenamefont {Risken}(1996)}]{risken1996fokker}%
  \BibitemOpen
  \bibfield  {author} {\bibinfo {author} {\bibfnamefont {H.}~\bibnamefont
  {Risken}},\ }\href@noop {} {\bibinfo {title} {The fokker-planck equation}}
  (\bibinfo {year} {1996})\BibitemShut {NoStop}%
\bibitem [{\citenamefont {Vinci}\ and\ \citenamefont
  {Mattia}(2024)}]{vinci2024escape}%
  \BibitemOpen
  \bibfield  {author} {\bibinfo {author} {\bibfnamefont {G.~V.}\ \bibnamefont
  {Vinci}}\ and\ \bibinfo {author} {\bibfnamefont {M.}~\bibnamefont {Mattia}},\
  }\bibfield  {title} {\bibinfo {title} {Escape time in bistable neuronal
  populations driven by colored synaptic noise},\ }\href@noop {} {\bibfield
  {journal} {\bibinfo  {journal} {arXiv preprint arXiv:2404.05391}\ } (\bibinfo
  {year} {2024})}\BibitemShut {NoStop}%
\bibitem [{\citenamefont {Shizgal}(2015)}]{shizgal2015spectral}%
  \BibitemOpen
  \bibfield  {author} {\bibinfo {author} {\bibfnamefont {B.}~\bibnamefont
  {Shizgal}},\ }\bibfield  {title} {\bibinfo {title} {Spectral methods in
  chemistry and physics},\ }\href@noop {} {\bibfield  {journal} {\bibinfo
  {journal} {Scientific Computation. Springer}\ } (\bibinfo {year}
  {2015})}\BibitemShut {NoStop}%
\bibitem [{\citenamefont {Williams}\ \emph {et~al.}(2015)\citenamefont
  {Williams}, \citenamefont {Kevrekidis},\ and\ \citenamefont
  {Rowley}}]{williams2015data}%
  \BibitemOpen
  \bibfield  {author} {\bibinfo {author} {\bibfnamefont {M.~O.}\ \bibnamefont
  {Williams}}, \bibinfo {author} {\bibfnamefont {I.~G.}\ \bibnamefont
  {Kevrekidis}},\ and\ \bibinfo {author} {\bibfnamefont {C.~W.}\ \bibnamefont
  {Rowley}},\ }\bibfield  {title} {\bibinfo {title} {A data--driven
  approximation of the koopman operator: Extending dynamic mode
  decomposition},\ }\href@noop {} {\bibfield  {journal} {\bibinfo  {journal}
  {Journal of Nonlinear Science}\ }\textbf {\bibinfo {volume} {25}},\ \bibinfo
  {pages} {1307} (\bibinfo {year} {2015})}\BibitemShut {NoStop}%
\bibitem [{\citenamefont {Colbrook}\ \emph {et~al.}(2023)\citenamefont
  {Colbrook}, \citenamefont {Ayton},\ and\ \citenamefont
  {Sz{\H{o}}ke}}]{colbrook2023residual}%
  \BibitemOpen
  \bibfield  {author} {\bibinfo {author} {\bibfnamefont {M.~J.}\ \bibnamefont
  {Colbrook}}, \bibinfo {author} {\bibfnamefont {L.~J.}\ \bibnamefont
  {Ayton}},\ and\ \bibinfo {author} {\bibfnamefont {M.}~\bibnamefont
  {Sz{\H{o}}ke}},\ }\bibfield  {title} {\bibinfo {title} {Residual dynamic mode
  decomposition: robust and verified koopmanism},\ }\href@noop {} {\bibfield
  {journal} {\bibinfo  {journal} {Journal of Fluid Mechanics}\ }\textbf
  {\bibinfo {volume} {955}},\ \bibinfo {pages} {A21} (\bibinfo {year}
  {2023})}\BibitemShut {NoStop}%
\bibitem [{\citenamefont {Colbrook}\ \emph {et~al.}(2019)\citenamefont
  {Colbrook}, \citenamefont {Roman},\ and\ \citenamefont
  {Hansen}}]{colbrook2019compute}%
  \BibitemOpen
  \bibfield  {author} {\bibinfo {author} {\bibfnamefont {M.~J.}\ \bibnamefont
  {Colbrook}}, \bibinfo {author} {\bibfnamefont {B.}~\bibnamefont {Roman}},\
  and\ \bibinfo {author} {\bibfnamefont {A.~C.}\ \bibnamefont {Hansen}},\
  }\bibfield  {title} {\bibinfo {title} {How to compute spectra with error
  control},\ }\href@noop {} {\bibfield  {journal} {\bibinfo  {journal}
  {Physical Review Letters}\ }\textbf {\bibinfo {volume} {122}},\ \bibinfo
  {pages} {250201} (\bibinfo {year} {2019})}\BibitemShut {NoStop}%
\bibitem [{\citenamefont {T{\"a}uber}(2014)}]{tauber2014critical}%
  \BibitemOpen
  \bibfield  {author} {\bibinfo {author} {\bibfnamefont {U.~C.}\ \bibnamefont
  {T{\"a}uber}},\ }\href@noop {} {\emph {\bibinfo {title} {Critical dynamics: a
  field theory approach to equilibrium and non-equilibrium scaling behavior}}}\
  (\bibinfo  {publisher} {Cambridge University Press},\ \bibinfo {year}
  {2014})\BibitemShut {NoStop}%
\bibitem [{Note1()}]{Note1}%
  \BibitemOpen
  \bibinfo {note} {$\mathinner {\langle {f(x)}\rangle }= \DOTSI \intop
  \ilimits@ _{\protect \mathcal {D}} f(x) p(x,t|x_0) p_s(x_0) dx
  dx_0$}\BibitemShut {NoStop}%
\bibitem [{\citenamefont {Brunton}\ \emph {et~al.}(2022)\citenamefont
  {Brunton}, \citenamefont {Budišić}, \citenamefont {Kaiser},\ and\
  \citenamefont {Kutz}}]{brunton2022modern}%
  \BibitemOpen
  \bibfield  {author} {\bibinfo {author} {\bibfnamefont {S.~L.}\ \bibnamefont
  {Brunton}}, \bibinfo {author} {\bibfnamefont {M.}~\bibnamefont {Budišić}},
  \bibinfo {author} {\bibfnamefont {E.}~\bibnamefont {Kaiser}},\ and\ \bibinfo
  {author} {\bibfnamefont {J.~N.}\ \bibnamefont {Kutz}},\ }\bibfield  {title}
  {\bibinfo {title} {Modern koopman theory for dynamical systems},\ }\href@noop
  {} {\bibfield  {journal} {\bibinfo  {journal} {SIAM Review}\ }\textbf
  {\bibinfo {volume} {64}},\ \bibinfo {pages} {229} (\bibinfo {year}
  {2022})}\BibitemShut {NoStop}%
\bibitem [{\citenamefont {Gardiner}\ and\ \citenamefont
  {Zoller}(2004)}]{gardiner2004quantum}%
  \BibitemOpen
  \bibfield  {author} {\bibinfo {author} {\bibfnamefont {C.}~\bibnamefont
  {Gardiner}}\ and\ \bibinfo {author} {\bibfnamefont {P.}~\bibnamefont
  {Zoller}},\ }\href@noop {} {\emph {\bibinfo {title} {Quantum noise: a
  handbook of Markovian and non-Markovian quantum stochastic methods with
  applications to quantum optics}}}\ (\bibinfo  {publisher} {Springer Science
  \& Business Media},\ \bibinfo {year} {2004})\BibitemShut {NoStop}%
\bibitem [{\citenamefont {Klus}\ \emph {et~al.}(2018)\citenamefont {Klus},
  \citenamefont {Bittracher}, \citenamefont {Schuster},\ and\ \citenamefont
  {Sch{\"u}tte}}]{klus2018kernel}%
  \BibitemOpen
  \bibfield  {author} {\bibinfo {author} {\bibfnamefont {S.}~\bibnamefont
  {Klus}}, \bibinfo {author} {\bibfnamefont {A.}~\bibnamefont {Bittracher}},
  \bibinfo {author} {\bibfnamefont {I.}~\bibnamefont {Schuster}},\ and\
  \bibinfo {author} {\bibfnamefont {C.}~\bibnamefont {Sch{\"u}tte}},\
  }\bibfield  {title} {\bibinfo {title} {A kernel-based approach to molecular
  conformation analysis},\ }\href@noop {} {\bibfield  {journal} {\bibinfo
  {journal} {The Journal of Chemical Physics}\ }\textbf {\bibinfo {volume}
  {149}} (\bibinfo {year} {2018})}\BibitemShut {NoStop}%
\bibitem [{\citenamefont {Ghanem}\ and\ \citenamefont
  {Spanos}(2003)}]{ghanem2003stochastic}%
  \BibitemOpen
  \bibfield  {author} {\bibinfo {author} {\bibfnamefont {R.~G.}\ \bibnamefont
  {Ghanem}}\ and\ \bibinfo {author} {\bibfnamefont {P.~D.}\ \bibnamefont
  {Spanos}},\ }\href@noop {} {\emph {\bibinfo {title} {Stochastic finite
  elements: a spectral approach}}}\ (\bibinfo  {publisher} {Courier
  Corporation},\ \bibinfo {year} {2003})\BibitemShut {NoStop}%
\bibitem [{\citenamefont {Vulpiani}\ \emph {et~al.}(2009)\citenamefont
  {Vulpiani}, \citenamefont {Cecconi},\ and\ \citenamefont
  {Cencini}}]{Vulpiani2009}%
  \BibitemOpen
  \bibfield  {author} {\bibinfo {author} {\bibfnamefont {A.}~\bibnamefont
  {Vulpiani}}, \bibinfo {author} {\bibfnamefont {F.}~\bibnamefont {Cecconi}},\
  and\ \bibinfo {author} {\bibfnamefont {M.}~\bibnamefont {Cencini}},\
  }\href@noop {} {\emph {\bibinfo {title} {Chaos: From simple models to complex
  systems}}},\ Series On Advances In Statistical Mechanics\ (\bibinfo
  {publisher} {World Scientific Publishing},\ \bibinfo {address} {Singapore,
  Singapore},\ \bibinfo {year} {2009})\BibitemShut {NoStop}%
\bibitem [{\citenamefont {Namdari}\ and\ \citenamefont
  {Li}(2019)}]{namdari2019review}%
  \BibitemOpen
  \bibfield  {author} {\bibinfo {author} {\bibfnamefont {A.}~\bibnamefont
  {Namdari}}\ and\ \bibinfo {author} {\bibfnamefont {Z.}~\bibnamefont {Li}},\
  }\bibfield  {title} {\bibinfo {title} {A review of entropy measures for
  uncertainty quantification of stochastic processes},\ }\href@noop {}
  {\bibfield  {journal} {\bibinfo  {journal} {Advances in Mechanical
  Engineering}\ }\textbf {\bibinfo {volume} {11}},\ \bibinfo {pages}
  {1687814019857350} (\bibinfo {year} {2019})}\BibitemShut {NoStop}%
\bibitem [{Note2()}]{Note2}%
  \BibitemOpen
  \bibinfo {note} {We verified that our results holds also for the case of
  small noise $\epsilon \sim 0.1$ but convergence requires almost double the
  number of dictionary functions, likely due the singular nature of the
  stationary measure}\BibitemShut {NoStop}%
\bibitem [{\citenamefont {Treves}(1993)}]{Treves1993}%
  \BibitemOpen
  \bibfield  {author} {\bibinfo {author} {\bibfnamefont {A.}~\bibnamefont
  {Treves}},\ }\bibfield  {title} {\bibinfo {title} {{Mean-field analysis of
  neuronal spike dynamics}},\ }\href@noop {} {\bibfield  {journal} {\bibinfo
  {journal} {Network}\ }\textbf {\bibinfo {volume} {4}},\ \bibinfo {pages}
  {259} (\bibinfo {year} {1993})}\BibitemShut {NoStop}%
\bibitem [{\citenamefont {Mattia}\ and\ \citenamefont {{Del
  Giudice}}(2002)}]{Mattia2002}%
  \BibitemOpen
  \bibfield  {author} {\bibinfo {author} {\bibfnamefont {M.}~\bibnamefont
  {Mattia}}\ and\ \bibinfo {author} {\bibfnamefont {P.}~\bibnamefont {{Del
  Giudice}}},\ }\bibfield  {title} {\bibinfo {title} {{Population dynamics of
  interacting spiking neurons}},\ }\href
  {https://doi.org/10.1103/PhysRevE.66.051917} {\bibfield  {journal} {\bibinfo
  {journal} {Phys. Rev. E}\ }\textbf {\bibinfo {volume} {66}},\ \bibinfo
  {pages} {051917} (\bibinfo {year} {2002})}\BibitemShut {NoStop}%
\bibitem [{\citenamefont {Jaeger}\ and\ \citenamefont
  {Haas}(2004)}]{Jaeger2004}%
  \BibitemOpen
  \bibfield  {author} {\bibinfo {author} {\bibfnamefont {H.}~\bibnamefont
  {Jaeger}}\ and\ \bibinfo {author} {\bibfnamefont {H.}~\bibnamefont {Haas}},\
  }\bibfield  {title} {\bibinfo {title} {{Harnessing nonlinearity: predicting
  chaotic systems and saving energy in wireless communication.}},\ }\href
  {https://doi.org/10.1126/science.1091277} {\bibfield  {journal} {\bibinfo
  {journal} {Science}\ }\textbf {\bibinfo {volume} {304}},\ \bibinfo {pages}
  {78} (\bibinfo {year} {2004})}\BibitemShut {NoStop}%
\bibitem [{\citenamefont {Pathak}\ \emph {et~al.}(2018)\citenamefont {Pathak},
  \citenamefont {Hunt}, \citenamefont {Girvan}, \citenamefont {Lu},\ and\
  \citenamefont {Ott}}]{pathak2018model}%
  \BibitemOpen
  \bibfield  {author} {\bibinfo {author} {\bibfnamefont {J.}~\bibnamefont
  {Pathak}}, \bibinfo {author} {\bibfnamefont {B.}~\bibnamefont {Hunt}},
  \bibinfo {author} {\bibfnamefont {M.}~\bibnamefont {Girvan}}, \bibinfo
  {author} {\bibfnamefont {Z.}~\bibnamefont {Lu}},\ and\ \bibinfo {author}
  {\bibfnamefont {E.}~\bibnamefont {Ott}},\ }\bibfield  {title} {\bibinfo
  {title} {Model-free prediction of large spatiotemporally chaotic systems from
  data: A reservoir computing approach},\ }\href@noop {} {\bibfield  {journal}
  {\bibinfo  {journal} {Phys. Rev. Lett.}\ }\textbf {\bibinfo {volume} {120}},\
  \bibinfo {pages} {024102} (\bibinfo {year} {2018})}\BibitemShut {NoStop}%
\bibitem [{\citenamefont {Di~Antonio}\ \emph {et~al.}(2024)\citenamefont
  {Di~Antonio}, \citenamefont {Gili}, \citenamefont {Gabrielli},\ and\
  \citenamefont {Mattia}}]{di2024linearizing}%
  \BibitemOpen
  \bibfield  {author} {\bibinfo {author} {\bibfnamefont {G.}~\bibnamefont
  {Di~Antonio}}, \bibinfo {author} {\bibfnamefont {T.}~\bibnamefont {Gili}},
  \bibinfo {author} {\bibfnamefont {A.}~\bibnamefont {Gabrielli}},\ and\
  \bibinfo {author} {\bibfnamefont {M.}~\bibnamefont {Mattia}},\ }\bibfield
  {title} {\bibinfo {title} {Linearizing and forecasting: a reservoir computing
  route to digital twins of the brain},\ }\href@noop {} {\bibfield  {journal}
  {\bibinfo  {journal} {bioRxiv}\ ,\ \bibinfo {pages} {2024}} (\bibinfo {year}
  {2024})}\BibitemShut {NoStop}%
\bibitem [{\citenamefont {Koopman}(1931)}]{koopman1931hamiltonian}%
  \BibitemOpen
  \bibfield  {author} {\bibinfo {author} {\bibfnamefont {B.~O.}\ \bibnamefont
  {Koopman}},\ }\bibfield  {title} {\bibinfo {title} {Hamiltonian systems and
  transformation in hilbert space},\ }\href@noop {} {\bibfield  {journal}
  {\bibinfo  {journal} {Proceedings of the National Academy of Sciences}\
  }\textbf {\bibinfo {volume} {17}},\ \bibinfo {pages} {315} (\bibinfo {year}
  {1931})}\BibitemShut {NoStop}%
\bibitem [{\citenamefont {Neumann}(1932)}]{neumann1932operatorenmethode}%
  \BibitemOpen
  \bibfield  {author} {\bibinfo {author} {\bibfnamefont {J.~v.}\ \bibnamefont
  {Neumann}},\ }\bibfield  {title} {\bibinfo {title} {Zur operatorenmethode in
  der klassischen mechanik},\ }\href@noop {} {\bibfield  {journal} {\bibinfo
  {journal} {Annals of Mathematics}\ }\textbf {\bibinfo {volume} {33}},\
  \bibinfo {pages} {587} (\bibinfo {year} {1932})}\BibitemShut {NoStop}%
\bibitem [{\citenamefont {Budi{\v{s}}i{\'c}}\ \emph {et~al.}(2012)\citenamefont
  {Budi{\v{s}}i{\'c}}, \citenamefont {Mohr},\ and\ \citenamefont
  {Mezi{\'c}}}]{budivsic2012applied}%
  \BibitemOpen
  \bibfield  {author} {\bibinfo {author} {\bibfnamefont {M.}~\bibnamefont
  {Budi{\v{s}}i{\'c}}}, \bibinfo {author} {\bibfnamefont {R.}~\bibnamefont
  {Mohr}},\ and\ \bibinfo {author} {\bibfnamefont {I.}~\bibnamefont
  {Mezi{\'c}}},\ }\bibfield  {title} {\bibinfo {title} {Applied koopmanism},\
  }\href@noop {} {\bibfield  {journal} {\bibinfo  {journal} {Chaos: An
  Interdisciplinary Journal of Nonlinear Science}\ }\textbf {\bibinfo {volume}
  {22}} (\bibinfo {year} {2012})}\BibitemShut {NoStop}%
\bibitem [{\citenamefont {Capone}\ \emph {et~al.}(2018)\citenamefont {Capone},
  \citenamefont {Gigante},\ and\ \citenamefont
  {Del~Giudice}}]{capone2018spontaneous}%
  \BibitemOpen
  \bibfield  {author} {\bibinfo {author} {\bibfnamefont {C.}~\bibnamefont
  {Capone}}, \bibinfo {author} {\bibfnamefont {G.}~\bibnamefont {Gigante}},\
  and\ \bibinfo {author} {\bibfnamefont {P.}~\bibnamefont {Del~Giudice}},\
  }\bibfield  {title} {\bibinfo {title} {Spontaneous activity emerging from an
  inferred network model captures complex spatio-temporal dynamics of spike
  data},\ }\href@noop {} {\bibfield  {journal} {\bibinfo  {journal} {Scientific
  reports}\ }\textbf {\bibinfo {volume} {8}},\ \bibinfo {pages} {17056}
  (\bibinfo {year} {2018})}\BibitemShut {NoStop}%
\bibitem [{\citenamefont {Lusch}\ \emph {et~al.}(2018)\citenamefont {Lusch},
  \citenamefont {Kutz},\ and\ \citenamefont {Brunton}}]{lusch2018}%
  \BibitemOpen
  \bibfield  {author} {\bibinfo {author} {\bibfnamefont {B.}~\bibnamefont
  {Lusch}}, \bibinfo {author} {\bibfnamefont {J.~N.}\ \bibnamefont {Kutz}},\
  and\ \bibinfo {author} {\bibfnamefont {S.~L.}\ \bibnamefont {Brunton}},\
  }\bibfield  {title} {\bibinfo {title} {Deep learning for universal linear
  embeddings of nonlinear dynamics},\ }\href@noop {} {\bibfield  {journal}
  {\bibinfo  {journal} {Nature communications}\ }\textbf {\bibinfo {volume}
  {9}},\ \bibinfo {pages} {4950} (\bibinfo {year} {2018})}\BibitemShut
  {NoStop}%
\bibitem [{\citenamefont {Alford-Lago}\ \emph {et~al.}(2022)\citenamefont
  {Alford-Lago}, \citenamefont {Curtis}, \citenamefont {Ihler},\ and\
  \citenamefont {Issan}}]{alford2022deep}%
  \BibitemOpen
  \bibfield  {author} {\bibinfo {author} {\bibfnamefont {D.~J.}\ \bibnamefont
  {Alford-Lago}}, \bibinfo {author} {\bibfnamefont {C.~W.}\ \bibnamefont
  {Curtis}}, \bibinfo {author} {\bibfnamefont {A.~T.}\ \bibnamefont {Ihler}},\
  and\ \bibinfo {author} {\bibfnamefont {O.}~\bibnamefont {Issan}},\ }\bibfield
   {title} {\bibinfo {title} {Deep learning enhanced dynamic mode
  decomposition},\ }\href@noop {} {\bibfield  {journal} {\bibinfo  {journal}
  {Chaos: An Interdisciplinary Journal of Nonlinear Science}\ }\textbf
  {\bibinfo {volume} {32}} (\bibinfo {year} {2022})}\BibitemShut {NoStop}%
\bibitem [{\citenamefont {Otto}\ and\ \citenamefont {Rowley}(2019)}]{Otto}%
  \BibitemOpen
  \bibfield  {author} {\bibinfo {author} {\bibfnamefont {S.~E.}\ \bibnamefont
  {Otto}}\ and\ \bibinfo {author} {\bibfnamefont {C.~W.}\ \bibnamefont
  {Rowley}},\ }\bibfield  {title} {\bibinfo {title} {Linearly recurrent
  autoencoder networks for learning dynamics},\ }\href
  {https://doi.org/10.1137/18M1177846} {\bibfield  {journal} {\bibinfo
  {journal} {SIAM Journal on Applied Dynamical Systems}\ }\textbf {\bibinfo
  {volume} {18}},\ \bibinfo {pages} {558} (\bibinfo {year} {2019})}\BibitemShut
  {NoStop}%
\end{thebibliography}%

\end{document}